\documentclass[prd,nofootinbib,amsmath,superscriptaddress,preprint]{revtex4-1}

\usepackage{graphicx}
\usepackage{color}
\usepackage[shortlabels]{enumitem}

\begin{document}
\title{Degenerate behavior in nonlinear vacuum electrodynamics}

\author{C. A. Escobar}
\email{carlos\_escobar@fisica.unam.mx}
\affiliation{Instituto de F\'\i sica, Universidad Nacional Aut\'onoma de M\'exico, Apdo.\ Postal 20-364, Ciudad de M\'exico 01000, M\'exico}
\author{R. Potting}
\email{rpotting@ualg.pt}
\affiliation{Departamento de F\'isica, Universidade do Algarve, FCT, 8005-139 Faro, Portugal}
\affiliation{CENTRA, Instituto Superior T\'ecnico, Universidade de Lisboa,
Avenida Rovisco Pais, Lisboa, Portugal}

\begin{abstract}
We study nonlinear vacuum electrodynamics in the first-order formulation
proposed by Pleba\'nski.
We analyze in detail the equations of motion,
and identify conditions for which a singularity can occur for the
time derivative of one of the field components.
The resulting degenerate behavior can give rise to a shock wave
with a reduction of the local number of degrees of freedom.
We use an example model to illustrate the occurrence of
superluminal propagation for field values approaching the singularity.
\end{abstract}


\maketitle

\section{Introduction}
\label{sec:introduction}

In 1934 Born and Infeld \cite{BornInfeld} proposed a nonlinear modification
of Maxwell theory with the objective to eliminate the infinite
self-energy of a point charge. 
The model follows from a Lorentz-invariant Lagrangian
and it gained renewed attention
when it was shown that it can be derived as an effective action from
quantized string theory \cite{FradkinTseytlin}.
Another nonlinear, Lorentz-invariant modification of Maxwell theory
was first derived from quantum electrodynamics by Heisenberg and Euler
\cite{EulerHeisenberg,Berestetskii}.
Pleba\'nski \cite{Plebanski} showed these models are two examples
of a large class of nonlinear electrodynamics (NLED) theories that are defined
by a gauge-invariant as well as Lorentz-invariant first-order Lagrangian.
The equations of motion of these models can be expressed like the usual 
Maxwell equations in material media,
where now the permitivities and permeabilities
are Lorentz-invariant functions of the electromagnetic field variables.
In recent decades models of nonlinear 
electrodynamics have been proposed in the context of QCD
(the Pagels-Tomboulis model \cite{Arodz}),
as arising from Kaluza-Klein compactification \cite{Lemos,Gibbons},
in order to explain accelerated expansion of the universe \cite{Novello},
in the context of galactic magnetic fields \cite{Campanelli}, etc.
There is also an extensive literature of solutions of nonlinear electrodynamics in a curved space background, such as black holes \cite{AyonBeato,Bronnikov}.

Wave propagation in the presence of background fields in NLED theories
has been studied by analyzing the so-called Fresnel equation,
which amounts to a dispersion relation for the wave vectors.
It can be derived by studying either the propagation of surfaces of discontinuities
\cite{nonlinear-propagation1,nonlinear-propagation1a,nonlinear-propagation2},
or by assuming an approximate a plane-wave ansatz \cite{Lammerzahl3}.
It is found that subluminal as well as superluminal propagation
is possible in general \cite{Lammerzahl,causal1},
and that birefringence effects can occur \cite{birefringence1,Lammerzahl2}.
In this work we will use a different approach and analyze
the NLED equations of motion directly for arbitrary potentials.

As we will show in this work, for certain NLED potentials there are hypersurfaces
in field space on which the equations of motion develop a singularity
in the sense that the time derivative of one of the field components
blows up.
Moreover, the equation of motion for this field component contains
an term which is multiplied by an undetermined coefficient,
and at the same time a certain (nonlinear) combination of the field variables
is forced to vanish.
This indicates a loss of two phase space degrees of freedom in the dynamical system, from the usual four (just like in Maxwell theory) to two.
We also consider a simple example in which we analyze wave propagation
in a background of nonzero fields,
and analyze what happens on points at which the background field lies
on a singular hypersurface.

This paper is organized as follows.
In section \ref{sec:first-order} we review the Pleba\'nski formulation
of NLED.
In section \ref{sec:degeneracy} we analyze the equations of motion,
with an emphasis on the degenerate behavior on certain hypersurfaces
in field space.
In section \ref{sec:linearized-eom} we analyze the linearized equations
of motion for fluctuations on a background of constant magnetic field,
for a particular NLED potential,
highlighting the behavior near the points on the degenerate hypersurface. 
Finally, we present our conclusions, as well as an outlook,
in section \ref{sec:conclusion}.

\section{First-order formulation of nonlinear electrodynamics}
\label{sec:first-order}
In this section we will review the first-order framework for
nonlinear electrodynamics that we will use in this work,
introducing notation and fixing our conventions.
Starting point is the action
\begin{equation}
S = \int d^4x\,\mathcal{L}
\label{action}
\end{equation}
in Minkowski space, with a Lagrangian density \cite{EscobarUrrutia}
\begin{equation}
\mathcal{L}= - P^{\mu\nu} \partial_\mu A_\nu - V(P,Q) - A_\mu J^\mu
\label{Lagrangian-density}
\end{equation}
that depends on the vector potential $A_\mu$ and on the antisymmetric tensor
$P^{\mu\nu}$,
which are treated as independent fields in \eqref{Lagrangian-density}.
The potential $V$ is taken to depend on $P^{\mu\nu}$ through the Lorentz scalars
\begin{equation}
P = \frac{1}{4}P_{\mu\nu} P^{\mu\nu}\>
\qquad\mbox{and}\qquad
Q = \frac{1}{4}P_{\mu\nu} \tilde P^{\mu\nu}\>
\label{P-Q}
\end{equation}
where the dual to $P^{\mu\nu}$ is defined by
\begin{equation}
\tilde P^{\mu\nu} = \frac{1}{2} \epsilon^{\mu\nu\rho\sigma} P_{\rho\sigma}\>.
\label{P-tilde}
\end{equation}
The Levi-Civita symbol is defined with the convention
$\epsilon^{0123} = -\epsilon_{0123} = 1$.
Note that in this work we assume the metric convention $(+,-,-,-)$ and use
natural, Heaviside-Lorentz units (with $c = \hbar = 1$).
From \eqref{P-tilde} we find the inverse relation
\begin{equation}
P^{\mu\nu} = -\frac{1}{2}\epsilon^{\mu\nu\rho\sigma} \tilde P_{\rho\sigma}\>.
\label{P-tilde-inverse}
\end{equation}
The last term in the Lagrangian density \eqref{Lagrangian-density}
defines a minimal coupling to the external current density $J^\mu$,
which is assumed to be conserved:
\begin{equation}
\partial_\mu J^\mu = 0\>.
\label{current-conservation}
\end{equation}
%
%
%
The equations of motion of \eqref{action} are
\begin{align}
\label{inhomogeneous-Maxwell}
\frac{\delta S}{\delta A_\mu} &= -\partial_\nu P^{\mu\nu} - J^\mu = 0 \>,\\
\frac{\delta S}{\delta P^{\mu\nu}} &=
- \frac12 F_{\mu\nu}
- \frac{1}{2}\left(V_P P_{\mu\nu} + V_Q \tilde P_{\mu\nu}\right) = 0 \>,
\label{Amu-equation}
\end{align}
where the lower indices on $V$ indicate the partial derivatives
$V_P = {\partial V}/{\partial P}$ and $V_Q = {\partial V}/{\partial Q}$,
and
\begin{equation}
F_{\mu\nu} \equiv \partial_\mu A_\nu -\partial_\nu A_\mu\>.
\label{F-A}
\end{equation}
Eq.\ \eqref{Amu-equation} becomes the constitutive relation
\begin{equation}
F_{\mu\nu} = -2\frac{\partial V}{\partial P^{\mu\nu}}
= - V_P P_{\mu\nu} - V_Q \tilde P_{\mu\nu}\>.
\label{constitutive-relation1}
\end{equation}
From definition \eqref{F-A} it follows that $F_{\mu\nu}$ satisfies
the consistency condition (Bianchi identity)
\begin{equation}
\partial_\mu \tilde{F}^{\mu\nu} = \frac{1}{2}\partial_\mu\epsilon^{\mu\nu\rho\sigma} F_{\rho\sigma} = 0\>.
\label{homogenous-Maxwell}
\end{equation}
The constitutive relation \eqref{constitutive-relation1} can be inverted
by considering $\mathcal{L}$ to be a function of $F^{\mu\nu}$ (as well as
$A_\mu$ and $J^\mu$). 
By Lorentz invariance, $\mathcal{L}$ should then be a function of the
invariants \cite{invariants}
\begin{equation}
F = \frac{1}{4}F_{\mu\nu} F^{\mu\nu}\>
\qquad\mbox{and}\qquad
G = \frac{1}{4}F_{\mu\nu} \tilde F^{\mu\nu}\>.
\label{F-G}
\end{equation}
Taking the variation of the action $S = \int \mathcal{L}(F,G)d^4x$
with respect to $F_{\mu\nu}$ and comparing it with the variation
of the action in its original form \eqref{action} it follows that
%
%
%
%
%
\begin{equation}
P^{\mu\nu} = -\mathcal{L}_F F^{\mu\nu} - \mathcal{L}_G \tilde F^{\mu\nu}
\label{constitutive-relation2}
\end{equation}
which expresses the inverse of the constitutive relation \eqref{constitutive-relation1}.

It will be useful in the following to express the above relations in terms
of the usual vector fields $\vec D$, $\vec E$, $\vec H$ and $\vec B$ by
defining
\begin{equation}
P^{\mu\nu} =
\begin{pmatrix}
0   & -D_x & -D_y & -D_z\\
D_x & 0    & -H_z & H_y\\
D_y & H_z  & 0    & -H_x\\
D_z & -H_y & H_x  & 0 
\end{pmatrix}
\label{Pmunu}
\end{equation}
and
\begin{equation}
F^{\mu\nu} =
\begin{pmatrix}
0    &  -E_x &  -E_y &  -E_z\\
E_x & 0    & -B_z & B_y\\
E_y & B_z  & 0    & -B_x\\
E_z & -B_y & B_x  & 0 
\end{pmatrix}\>.
\end{equation}
%
%
%
The invariants $P$, $Q$, $F$ and $G$ can then be written as
\begin{align}
P &= \frac{1}{2}(\vec{H}^2 - \vec{D}^2)
\qquad \mbox{and} \qquad
Q = -\vec H \cdot \vec D\>,
\label{PQ-HD} \\
F &= \frac{1}{2}(\vec B^2 - \vec E^2)
\qquad \mbox{and} \qquad
G = -\vec B \cdot \vec E\,.
\label{FG-BE}
\end{align}
while the constitutive relations \eqref{constitutive-relation1} and
\eqref{constitutive-relation2} can be expressed in matrix form:
\begin{equation}
\begin{pmatrix}\vec E \\ \vec B \end{pmatrix} =
\begin{pmatrix}-V_P & -V_Q \\ V_Q & -V_P \end{pmatrix}
\begin{pmatrix}\vec D \\ \vec H \end{pmatrix}
\label{constitutive-relations3a}
\end{equation}
and
\begin{equation}
\begin{pmatrix}\vec D \\ \vec H \end{pmatrix} =
\begin{pmatrix}-\mathcal{L}_F & -\mathcal{L}_G \\ \mathcal{L}_G & -\mathcal{L}_F \end{pmatrix}
\begin{pmatrix}\vec E \\ \vec B \end{pmatrix}\>.
\label{constitutive-relations3b}
\end{equation}
%
Equations \eqref{inhomogeneous-Maxwell} and \eqref{homogenous-Maxwell}
then take the familiar form of the Maxwell equations in a material medium
\begin{align}
\vec\nabla \cdot \vec D &= J^0 \>,\label{Gauss-D}\\
\vec\nabla \times \vec H -\frac{\partial\vec D}{\partial t} &= \vec J\>, \label{Ampere}\\
\vec\nabla \cdot \vec B &= 0\>, \label{Gauss-B}\\
\vec\nabla \times \vec E + \frac{\partial\vec B}{\partial t} &= 0\>. \label{Faraday}
\end{align}
The relations \eqref{constitutive-relations3a} and
\eqref{constitutive-relations3b}  yield, by consistency,
\begin{equation}
\begin{pmatrix}-V_P & -V_Q \\ V_Q & -V_P \end{pmatrix}
=
\begin{pmatrix}-\mathcal{L}_F & -\mathcal{L}_G \\ \mathcal{L}_G & -\mathcal{L}_F \end{pmatrix}^{-1}
\end{equation}
from which it follows that
\begin{equation}
\mathcal{L}_F = \frac{V_P}{V_P^2 + V_Q^2}
\qquad\mbox{and}\qquad
\mathcal{L}_G = \frac{-V_Q}{V_P^2 + V_Q^2}\>.
\label{potential-relations}
\end{equation}
For the energy-momentum tensor we get from Eq.\ \eqref{Lagrangian-density}
\begin{align}
T_{\mu\nu} &= P_\mu{}^\lambda F_{\nu\lambda} - \eta_{\mu\nu}\mathcal{L}\nonumber\\
&= V_P P_\mu{}^\lambda P_{\nu\lambda} - V_Q P_\mu{}^\lambda \tilde P_{\nu\lambda}
+ 2\eta_{\mu\nu}\bigl(V_P P + V_Q Q\bigr)
\label{energy-momentum}
\end{align}
where in the second identity we used the constitutive relation \eqref{constitutive-relation1}.
For the energy density we obtain
\begin{equation}
U = T_{00} = V - V_P |\vec{H}|^2 - V_Q Q
\label{energy-density}
\end{equation}
where we made use of the relations \eqref{PQ-HD} and the explicit form
\eqref{Pmunu} of $P_{\mu\nu}$.
\\

It is instructive to consider a few well-known special cases.

The usual Maxwell equations in vacuum follow by taking
$\mathcal{L}(F,G) = -F$, so that  $\mathcal{L}_F = -1$ and $\mathcal{L}_G = 0$.
It then follows that $P^{\mu\nu} = -\mathcal{L}_F F^{\mu\nu} = F^{\mu\nu}$,
so that the constitutive relations are trivial:
$\vec D = \vec E$ and $\vec H = \vec B$.
From relations \eqref{potential-relations} we find $V_P = -1$, $V_Q = 0$,
so that $V(P,Q) = -P$.

Another example is given by the Born-Infeld action,
which is defined by the potential
\begin{equation}
V(P,Q) = b^2\sqrt{1 - \dfrac{2P}{b^2} - \dfrac{Q^2}{b^4}} - b^2 \>.
\label{BI_V}
\end{equation}
where the Born-Infeld parameter $b$ has mass dimension 2.
From its partial derivatives $V_P$ and $V_Q$
\begin{equation}
\label{BI_VPQ}
V_P = \dfrac{-1 }{\sqrt{1 - \dfrac{2P}{b^2} - \dfrac{Q^2}{b^4}}}
\qquad\mbox{and}\qquad
V_Q = \dfrac{\dfrac{Q}{b^2}}{\sqrt{1 - \dfrac{2P}{b^2} - \dfrac{Q^2}{b^4}}}
\end{equation}
we can find expressions for $\mathcal{L}_F$ and $\mathcal{L}_G$
in terms of the invariants $P$ and $Q$
by using relations \eqref{potential-relations}.
It is straightforward to obtain explicit relations for $P$ and $Q$ in terms
of $F$ and $G$ (and vice-versa) by using the constitutive relation \eqref{constitutive-relation1} together with the expressions \eqref{BI_VPQ}.
It follows that
\begin{equation}
\mathcal{L}_F = \dfrac{-1}{\sqrt{1 + \dfrac{2F}{b^2} - \dfrac{G^2}{b^4}}}
\qquad\mbox{and}\qquad
\mathcal{L}_G = \dfrac{\dfrac{G}{b^2}}{\sqrt{1 + \dfrac{2F}{b^2} - \dfrac{G^2}{b^4}}}\>.
\label{BI-LFG}
\end{equation}
which can be integrated to yield the explicit form of the Born-Infeld Lagrangian
\begin{equation}
\mathcal{L} = b^2 - b^2\sqrt{1 + \frac{2F}{b^2} - \frac{G^2}{b^4}}
\label{BI_L}
\end{equation}
in terms of the invariants $F$ and $G$.
For a detailed treatment of the Born-Infeld model in the context of
the first-order formalism see \cite{Plebanski}.
\\

While for the Born-Infeld potential 
it is relatively straightforward to obtain the dual expression
for the associated Lagrangian as a function of the invariants $F$ and $G$
in explicit form, 
this is not always the case for general Pleba\'nski models.
In fact, the relations for $P$ and $Q$ in terms
of $F$ and $G$ are not even always invertible,
giving rise to branch points \cite{Bronnikov}.
In this case the physics described by the dual Lagrangian formulation only 
coincides with the original one for a restricted set of field values.
We will see that this issue is particularly important for the range of field
values that give rise to the singular behavior we will study below.
\\

As we will use the fields $\vec{D}$ and $\vec{H}$
to describe the dynamics of the theory, we need their time development.
The time development of the $\vec{D}$ field is directly determined
by Amp\`ere's law \eqref{Ampere}.
In order to extract the time development of $\vec{H}$ from Faraday's
law \eqref{Faraday}, it is necessary to use the constitutive
relations \eqref{constitutive-relations3a}.
They depend on the quantities $V_P$ and $V_Q$,
which in turn are functions of $\vec{H}$ and $\vec{D}$.
This makes determining the time development of $\vec{H}$ less than trivial.
Explicitly, from Eqs.\ \eqref{Faraday} and \eqref{constitutive-relations3a}
one finds
\begin{equation}
\frac{\partial}{\partial t}(V_Q\vec{D} - V_P\vec{H}) =
\vec{\nabla}\times(V_P\vec{D} + V_Q\vec{H})\>.
\label{Faraday_constitutive}
\end{equation}
For the time derivatives of $V_P$ and $V_Q$ we have
\begin{align}
\frac{\partial V_P}{\partial t}
&= V_{PP}\frac{\partial P}{\partial t} + V_{PQ}\frac{\partial Q}{\partial t}
\nonumber\\
&= (V_{PP}\vec{H} - V_{PQ}\vec{D})\cdot\frac{\partial \vec{H}}{\partial t}
- (V_{PP}\vec{D} + V_{PQ}\vec{H})\cdot\frac{\partial \vec{D}}{\partial t}
\label{dV_Pdt}
\end{align}
and
\begin{align}
\frac{\partial V_Q}{\partial t}
&= V_{PQ}\frac{\partial P}{\partial t} + V_{QQ}\frac{\partial Q}{\partial t}
\nonumber\\
&= (V_{PQ}\vec{H} - V_{QQ}\vec{D})\cdot\frac{\partial \vec{H}}{\partial t}
- (V_{PQ}\vec{D} + V_{QQ}\vec{H})\cdot\frac{\partial \vec{D}}{\partial t}\>.
\label{dV_Qdt}
\end{align}
One can now work out the time derivatives on the left-hand side of
Eq.\ \eqref{Faraday_constitutive},
substitute the expressions \eqref{dV_Pdt} and \eqref{dV_Qdt}.
The resulting equation can be written as
\begin{align}
\frac{\partial \vec{H}}{\partial t} = \frac{1}{V_P}\Bigl[&
V_Q\frac{\partial \vec{D}}{\partial t} - \vec{\nabla}\times(V_P\vec{D} + V_Q\vec{H})\nonumber\\
&{}- \Bigl((V_{PP}\vec{H} - V_{PQ}\vec{D})\cdot\frac{\partial\vec{H}}{\partial t} -
(V_{PP}\vec{D} + V_{PQ}\vec{H})\cdot\frac{\partial\vec{D}}{\partial t}\Bigl)
\,\vec{H} \nonumber\\
&{}+ \Bigl((V_{PQ}\vec{H} - V_{QQ}\vec{D})\cdot\frac{\partial\vec{H}}{\partial t} -
(V_{PQ}\vec{D} + V_{QQ}\vec{H})\cdot\frac{\partial\vec{D}}{\partial t}\Bigl)
\,\vec{D} \Bigr]
\label{dHdt}
\end{align}
The scalar products $\vec{D}\cdot(\partial\vec{H}/\partial t)$
and $\vec{H}\cdot(\partial\vec{H}/\partial t)$ on the right-hand side
of the equation can be obtained by taking the scalar product of
Eq.\ \eqref{dHdt} with $\vec{D}$ and with $\vec{H}$, respectively.
We then obtain the following expressions for $\vec{H}\cdot(\partial\vec{H}/\partial t)$
and $\vec{D}\cdot(\partial\vec{H}/\partial t)$:
\begin{align}
\label{H-dot-dHdt}
\vec{H}\cdot\frac{\partial \vec{H}}{\partial t} &= -S^{-1} \bigl(
(V_P + QV_{PQ} + D^2V_{QQ})A_1 + (H^2V_{PQ} + QV_{QQ})A_2\bigr)\\
\vec{D}\cdot\frac{\partial \vec{H}}{\partial t} &= -S^{-1} \bigl(
(QV_{PP} + D^2V_{PQ})A_1 + (V_P + H^2V_{PP} + QV_{PQ})A_2\bigr) \>,
\label{D-dot-dHdt}
\end{align}
with
\begin{align}
\label{A}
A_1 &= \vec{H}\cdot\vec{\nabla}\times(V_P\vec{D} + V_Q\vec{H})\nonumber\\
&\qquad{} -
\left((V_Q + H^2V_{PQ} + QV_{QQ})\vec{H} + (H^2V_{PP} + QV_{PQ})\vec{D}\right)
\cdot\frac{\partial \vec{D}}{\partial t}\\
\label{B}
A_2 &= \vec{D}\cdot\vec{\nabla}\times(V_P\vec{D} + V_Q\vec{H}) \nonumber\\
&\qquad{} +
\left((QV_{PQ} + D^2V_{QQ})\vec{H} + (-V_Q + QV_{PP} + D^2V_{PQ})\vec{D}\right)
\cdot\frac{\partial \vec{D}}{\partial t}\>
\end{align}
and
\begin{equation}
S = \bigl(V_{PP}V_{QQ}-V_{PQ}^2\bigr)\bigl(H^2D^2 - Q^2\bigr)
 + V_P\bigl(H^2 V_{PP} + 2 Q V_{PQ} + D^2 V_{QQ}\bigr) + V_P^2\>,
\label{S}
\end{equation}
where we introduced the notation $D = |\vec{D}|$ and $H = |\vec{H}|$.

It is evident from the (quite complicated) Eq.\ \eqref{dHdt} that
the time derivative of $\vec{H}$ is well defined and non-singular
as long as $V_P \ne 0$ and $S \ne 0$.
However, when $V_P \to 0$ or $S \to 0$, $\partial\vec{H} / \partial t$
can diverge.

\section{Global dynamics and degeneracy surfaces}
\label{sec:degeneracy}

We already saw in section \ref{sec:first-order} that degenerate behavior
can occur if either of the conditions
\begin{align}
\label{S=0}
S &= 0 \\
V_P  &= 0
\label{V_P=0}
\end{align}
(where $S$ is given by Eq.\ \eqref{S}) is satisfied.
In particular,
the time derivative of $\vec{H}$ given by Eq.\ \eqref{dHdt}
can diverge if either $V_P$ or $S$ tends to zero.

Let us first consider the degenerate dynamics of the $\vec{H}$ field
close to the hypersurface \underline{$S = 0$, $V_P \ne 0$}.
For simplicity, we will take the potential $V$
only to depend on $P$, not on $Q$.
Note that in that case $S = V_P(V_P + H^2 V_{PP})$ and thus the hypersurface
$S = 0$ then becomes equivalent to the condition $V_P + H^2V_{PP} = 0$.

Now as long as $V_P + H^2V_{PP}$ is small but nonzero,
we can use expression \eqref{dHdt}
which simplifies substantially because we can put to zero all terms
involving $V_Q$, $V_{PQ}$ and $V_{QQ}$.
It then follows that 
\begin{equation}
\dot{\vec{H}} = \dfrac{1}{V_P}\left(-\vec{\nabla}\times(V_P\vec{D})
+ \dfrac{NV_{PP}\vec{H}}{V_P + H^2V_{PP}}\right)\>,
\label{dot-H}
\end{equation}
where we defined
\begin{equation}
N \equiv \vec{H}\cdot(\vec{\nabla}\times V_P\vec{D}) + V_P\vec{D}\cdot(\vec{\nabla}\times\vec{H} - \vec{J})\>.
\label{N}
\end{equation}
Note that the right-hand-side of \eqref{dot-H} is (potentially) singular
in the limit $V_P + H^2V_{PP} \to 0$.

However, if $V_P + H^2V_{PP}$ is exactly zero,
Eqs.\ \eqref{Faraday_constitutive} and \eqref{Ampere} imply that
\begin{equation}
\dot{\vec{H}} = -\dfrac{\vec{\nabla}\times(V_P\vec{D})}{V_P} + \beta\vec{H}
\label{dot-H_degenerate}
\end{equation}
where the parameter $\beta$ is free,
together with the condition
\begin{equation}
N = 0\>.
\label{condition-N}
\end{equation}
Thus we see that on the $V_P + H^2V_{PP} = 0$ surface the time derivative
of the longitudinal component of $\vec{B}$ (that is, fluctuations of
$\vec{B}$ that modify its modulus, not its direction) are undetermined
and thus could be arbitrarily large.
Condition \eqref{condition-N} represents an extra constraint that is forced
on the field configuration.
The way this can be interpreted is that 
the freedom in the value of $\beta$ allows for the modulus of $\vec{H}$
to be adapted
such that condition \eqref{condition-N} is satisfied.

Now suppose that we are in a region of spacetime with non-constant
values of $V_P + H^2V_{PP}$, including the value zero.
The points in which $V_P + H^2V_{PP} = 0$ can then be expected to form a
two-dimensional surface $\Sigma$ in space.
We see then from Eq.\ \eqref{dot-H} that,
if on those points $N$ takes nonzero values,
the time derivatives of $\vec{H}$ will then diverge as we approach $\Sigma$.
To understand better what happens on the points close to $\Sigma$,
it is instructive to evaluate the time derivative of the \textit{square} of
$V_P + H^2V_{PP}$. One finds
\begin{align}
\frac{1}{2}\frac{\partial}{\partial t}\left((V_P + H^2V_{PP})^2\right) &=
-(3V_{PP} + H^2V_{PPP})N \nonumber\\
&\qquad{} + 2(V_P + H^2V_{PP})V_{PP}\vec{D}\cdot(\vec{\nabla}\times\vec{H} - \vec{J}\,)
\label{derivative-square}
\end{align}
where we assumed that $V_P + H^2V_{PP} \ne 0$.
In the limit $V_P + H^2V_{PP} \to 0$ the second term tends to zero,
leaving us with a finite, generally nonzero, value (given by the first term).

There are three possibilities for the values of $V_P + H^2V_{PP}$ on points close
to $\Sigma$.
\begin{itemize}
\item \underline{If $(3V_{PP} + H^2V_{PPP})N < 0$},
the absolute value of $V_P + H^2V_{PP}$ will grow in time.
In other words $V_P + H^2V_{PP}$ will be driven away from zero, in opposite
directions depending on the value (positive or negative).
\item \underline{If $(3V_{PP} + H^2V_{PPP})N > 0$},
the absolute value of $V_P + H^2V_{PP}$ will decrease in time.
In fact, as they approach zero, their time derivative will diverge,
indicating a possible discontinuity in the field values.
Thus $V_P + H^2V_{PP}$ will be driven to zero,
whether its value is positive or negative.
There even appears to be a paradox
if the time derivative of the non-negative quantity
$(V_P + H^2V_{PP})^2$ stays negative in the limit  $V_P + H^2V_{PP} \to 0$.
Note, however, that as soon as $V_P + H^2V_{PP}$ becomes zero the time derivative
of $\vec{H}$ is no longer given by Eq.\ \eqref{dot-H},
but by Eq.\ \eqref{dot-H_degenerate}.
It is then easy to check that 
$\frac{\partial}{\partial t}\bigl((V_P + H^2V_{PP})^2\bigr) = 0$,
eliminating the apparent inconsistency encountered above.
Moreover, note that, by condition \eqref{condition-N},
the quantity $N$ is forced to become zero.
\item \underline{The case  $(3V_{PP} + H^2V_{PPP})N = 0$}
interpolates between the two above situations.
\end{itemize}

Thus we find the following picture.
Whenever there is a spatial surface $\Sigma$ on which $V_P + H^2V_{PP}$ vanishes,
while the quantity $(3V_{PP} + H^2V_{PPP})N$ is positive,
points on which $V_P + H^2V_{PP}$ is close to zero are driven to zero.
Thus the surface $\Sigma$ can be expected to grow to a region of finite
volume, and will continue to grow as long as the value of the quantity
$(3V_{PP} + H^2V_{PPP})N$ on the points bordering (but outside) $\Sigma$
is positive.
Inside the region $\Sigma$ there are two conditions that have
to be satisfied.
First of all, the quantity $N$ has to remain zero.
Therefore, its time derivative has to vanish.
This yields the condition
\begin{align}
\dot{N} &= 2\vec{\nabla}\beta \cdot\bigl(\vec{E}\times\vec{H}\bigr)
-\beta\vec{J}\cdot\vec{E} + V_P\bigl|\vec{\nabla}\times\vec{H} - \vec{J}\bigr|^2
- \frac{1}{V_P}\bigl|\vec{\nabla}\times\vec{E}\bigr|^2 \nonumber\\
&\qquad{}- \vec{E}\cdot\left(\vec{\nabla}\times\bigl(\frac{1}{V_P}\vec{\nabla}\times\vec{E}\bigr) - \dot{\vec{J}}\right)
+ \vec{H}\cdot\vec{\nabla}\times\left(V_P\bigl(\vec{\nabla}\times\vec{H}
- \vec{J}\bigr)\right)\nonumber\\
&= 0\>.
\label{dot-N}
\end{align}
Thus the parameter $\beta$ has to satisfy a first-order
partial differential equation.
The latter allows for boundary conditions on
$\beta$ to be chosen on a two-dimensional surface, which can be
taken to coincide with the boundary of $\Sigma$.
A second condition is that the value of $V_P + H^2V_{PP}$ has to vanish.
For its time derivative we find
\begin{equation}
\frac{\partial}{\partial t}(V_P + H^2V_{PP}) =
2V_{PP} \vec{D}\cdot(\vec{\nabla}\times\vec{H} - \vec{J})
+ \beta H^2(3V_{PP} + H^2V_{PPP})\>.
\label{derivative}
\end{equation}
As long as the system remains on $\Sigma$,
this has to vanish for consistency, fixing the value of $\beta$.
Conditions \eqref{derivative} and \eqref{dot-N} can be satisfied
jointly at least on the two-dimensional boundary of $\Sigma$.
For the points inside the volume of $\Sigma$ to remain on the degenerate
surface it is necessary that conditions \eqref{derivative} and \eqref{dot-N}
remain both satisfied.
If not, these points may be forced off,
after which they will either fall back on the surface,
or are driven away, depending on the sign of the expression
on the right-hand side of Eq.\ \eqref{derivative-square}.

The dynamical surface defined this way behaves very much like a shock wave
separating two regions on which the expression $V_{P} + H^2V_{PP}$ is
either zero on nonzero.
The shock wave moves toward the region in which $V_{P} + H^2V_{PP}$ is nonzero
whenever the quantity $(3V_{PP} + H^2V_{PPP})N$ is positive,
thereby increasing the size of the region $\Sigma$,
while the opposite happens when $(3V_{PP} + H^2V_{PPP})N$ is negative.

That such shock waves turn up should perhaps not come as a surprise.
It is well known that shock waves can
be produced out of a continuous initial state in nonlinear electrodynamics
\cite{LutzkyToll},
except in certain ``exceptional'' models including
Born-Infeld electrodynamics \cite{Boillat-1970,Gibbons}.
At a shock wave the characteristics associated with the
partial differential equations describing the field dynamics intersect,
at which point the field equations cease to determine uniquely the
time development of the associated field components.
A similar failure of the equations of motions to uniquely determine the
time development of the fields is evident from Eq.\ \eqref{dot-H_degenerate}
due to the presence of the free parameter $\beta$.
It occurs together with the appearance of the extra constraint \eqref{condition-N},
reducing the number of local phase space degrees of freedom from
four to two.

As an aside, we note that the points on which $V_P + H^2V_{PP} = 0$ have
a particular significance.
It follows from the constitutive relation $\vec{B} = -V_P\vec{H}$
and the fact that $V_P$ only depends on $P = (H^2 - D^2)/2$ that
$B = |\vec{B}|$ is a function of $H$ and $D$.
One readily verifies that
\begin{equation}
\frac{\partial B}{\partial H} = V_P + H^2V_{PP}\>,
\end{equation}
so that the condition $V_P + H^2 V_{PP} = 0$ corresponds to
stationary points of the modulus of $\vec{B}$ as a function of $H$,
taking $D$ constant.
In the cases of interest $B$ actually has a local minimum.
\\

Next, we consider the degenerate dynamics of the $\vec{H}$ field
close to the hypersurface \underline{$S \ne 0$, $V_P = 0$}.
Here we will take the potential to be an arbitrary function
of the quantities $P$ and $Q$.
For the time derivative of $V_P$ it follows that
\begin{equation}
\frac{\partial V_P}{\partial t} = V_{PP}\left(\vec{H}\cdot\frac{\partial\vec{H}}{\partial t} - \vec{D}\cdot\frac{\partial\vec{D}}{\partial t}\right)
- V_{PQ}\left(\vec{D}\cdot\frac{\partial\vec{H}}{\partial t} + \vec{H}\cdot\frac{\partial\vec{D}}{\partial t}\right)\>.
\label{dV_P-dt}
\end{equation}
Using Eqs.\ \eqref{H-dot-dHdt}, \eqref{D-dot-dHdt}, \eqref{A} and
\eqref{B}, we can evaluate the right-hand side of Eq.\ \eqref{dV_P-dt}.
Unlike the situation close to the hypersurface $S = 0$ described above,
it turns out that in the limit $V_P \to 0$, the time derivative of $V_P$
is finite:
\begin{equation}
\lim_{V_P \to 0}\frac{\partial V_P}{\partial t} = \frac{-Q}{H^2 D^2 - Q^2}\vec{\nabla}V_Q\cdot(\vec{H}\times\vec{D})\>.
\label{dV_P-dt}
\end{equation}
Therefore
\begin{equation}
\lim_{V_P\to0}\frac{\partial \bigl((V_P)^2\bigr)}{\partial t} =
\lim_{V_P\to0} 2V_P\frac{\partial V_P}{\partial t} = 0
\end{equation}
and thus the dynamics does not suffer the same kind of singular behaviour
at the $V_P = 0$ surface as we encountered for the case $S = 0$.
The only possible exception can occur whenever the quantity $H^2D^2 - Q^2$
turns equal to zero (which happens when $\vec{H}$ and $\vec{D}$ are
parallel). 
However we will not investigate this possibility in detail in this work.
In particular, $V_P$ just evolves regularly when passing through the
surface, as described by Eq.\ \eqref{dV_P-dt}.

\section{Analysis of the linearized equations of motion away from the vacuum}
\label{sec:linearized-eom}

In this section we will consider the case of a potential $V(P)$ which,
for simplicity, is taken independent of $Q$.
The equations of motion then reduce to
\begin{align}
\label{eq_motion1}
\partial_t \vec{D} &= \vec{\nabla}\times\vec{H} \\
\partial_t(V_P\vec{H}) &= -\vec{\nabla}\times(V_P\vec{D})\>.
\label{eq_motion2}
\end{align}
We now write the fields $\vec{H}$ and $\vec{D}$ as fluctuations around
a constant field configuration $P_0 = \{\vec{H}_0,\vec{D}_0\}$,
where we will choose $\vec{D}_0 = 0$.
Thus
\begin{align}
\vec{H} &= \vec{H}_0 + \vec{h} \\
\vec{D} &= \vec{d}\>.
\end{align}

The equations of motion \eqref{eq_motion1} and \eqref{eq_motion2} then
become, up to linear order
\begin{align}
\label{eq_motion1-linear}
\partial_t \vec{d} &= \vec{\nabla}\times\vec{h} \\
\label{eq_motion2x-linear}
(V_P + H_0^2V_{PP})\partial_t h_\| &= -V_P(\vec{\nabla}\times\vec{d})_\|\\
\partial_t \vec{h}_\bot &= -(\vec{\nabla}\times\vec{d})_\bot
\label{eq_motion2perp-linear}
\end{align}
(here we have dropped the dependence of $V_P$ and $V_{PP}$ on $P_0$).
The $\|$ and $\bot$ indices indicate the components parallel and
perpendicular to $\vec{H}_0$.

It is easy to check that there is a massless transverse mode with
the polarization of the $\vec{h}_\bot$ vector perpendicular to $\vec{k}$
(as well as to $\vec{H}_0$) and $\vec{d}$ perpendicular to $\vec{h}_\bot$
and $\vec{k}$ that has the usual phase (or group) velocity 1.

\begin{figure}
  \centering
    \includegraphics{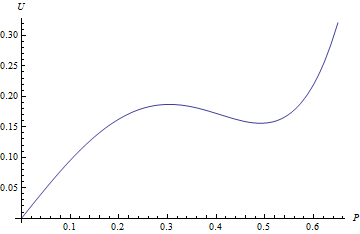}
  \caption{Form of the energy density as a function of
  $P = |\vec{H}_0|^2/2$ for the potential $V(P) = -P + P^3 - P^5$.}
  \label{fig:Heff}
\end{figure}

\begin{figure}
  \centering
    \includegraphics{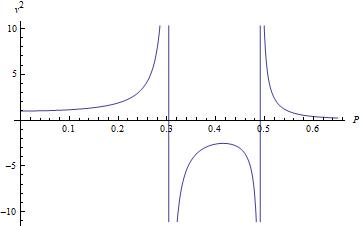}
  \caption{The dependence of $v^2 = \frac{V_P}{V_P + |\vec{H}_0|^2V_{PP}}$
  on $P = |\vec{H}_0|^2/2$ for the potential $V(P) = -P + P^3 - P^5$.}
  \label{fig:v2}
\end{figure}

To analyze the remaining mode, we choose Cartesian coordinates
such that $\vec{H}_0 = H_0 \vec{e}_x$.
Eqs.\ \eqref{eq_motion1-linear} and \eqref{eq_motion2x-linear} then
yield
\begin{equation}
\partial_t^2 h_x = \left(\partial_x^2 + \frac{V_P}{V_P + |\vec{H}_0|^2V_{PP}}(\partial_y^2 + \partial_z^2)\right)h_x\>.
\label{eq_hx}
\end{equation}
(where $h_x = h_\|$).
Let us suppose first, for simplicity, that $h_x$ is independent of $x$.
It then follows that, if $1 + |\vec{H}_0|^2\frac{V_{PP}}{V_P} > 0$,
this represents a massless mode with phase (as well as group) velocity
equal to $v_{ph} = \sqrt{\frac{V_P}{V_P + |\vec{H}_0|^2V_{PP}}}$.
However, if $1 + |\vec{H}_0|^2\frac{V_{PP}}{V_P} < 0$,
there is no plane-wave type propagation. 
In Fig.\ \ref{fig:Heff} we plot, as an example,
the form of the energy density \eqref{energy-density} for the potential
\begin{equation}
V(P) = -P + P^3 - P^5
\end{equation}
and, in Fig.\ \ref{fig:v2},
the corresponding dependence of the square of the propagation velocity.
We see that latter is negative in the interval between the local maximum
and minimum of the energy density.
Moreover, when approaching the local minimum from the right,
the phase/group velocity tends to infinity!

At the points where $V_P + |\vec{H}_0|^2 V_{PP} = 0$,
the equation of motion degenerates into the constraint equation
$(\partial_y^2 + \partial_z^2) h_x = 0$.
Thus, the fluctuations parallel to $\vec{H}_0$ stop being physical,
together with the appearance of a constraint.
This parallels the appearance of the free parameter $\beta$
in the equation of motion \eqref{dot-H_degenerate} for the $\vec{H}$ field
on the degenerate surface together with the appearance of the condition
\eqref{condition-N}.

\section{Discussion}
\label{sec:conclusion}

In this work, we considered a large class of nonlinear vacuum electrodynamics
models using a first-order approach introduced by Pleba\'nski \cite{Plebanski}.
We analyzed in detail the equations of motion, focusing on situations
in which the equations of motion develop a singularity.
In particular, this can happen for the equation of motion for the $\vec{H}$
field.
At the space-time points close to a singularity,
the time derivative of $\vec{H}$ can diverge,
while on the space-time points with the singularity,
the time derivative of $\vec{H}$ (more precisely, its modulus)
acquires an indeterminacy.
When this happens, the equations of motion imply also that
an extra constraint is turned on,
thus reducing the number of local degrees of freedom.

As we have shown, the dynamics of the region (surface) of degeneracy
has the behavior of a shock wave.
Shock waves have been shown to appear in general models of nonlinear
electrodynamics in the context of the propagation of linear disturbances
in an electromagnetic field background due to the possibility
of the formation of caustics \cite{Boillat-1970,Gibbons}.
They have also been shown to arise in a study of
the Euler-Heisenberg Lagrangian \cite{Zheleznyakov}.
Of course it is no surprise that we have encountered them as well in our
rather different approach.

In section \ref{sec:linearized-eom}  we analyzed,
as an illustration, the linearized field fluctuations in a constant
background magnetic field for a particular Pleba\'nski model. 
We showed that one of the propagation modes is superluminal for a range
of background field strengths,
while a singularity is encountered in the equation of motion as the 
field approaches its critical value.
At that point, the equation of motion turns into a constraint equation,
confirming the breakdown in the number of degrees of freedom.
From this example it is reasonable to expect also for general
Pleba\'nski models, that as the singularity is approached, there is a
range of field values for which either the phase velocity of perturbations
is larger than one (the speed of light),
or the field equation does not permit wave-like solutions.

In the literature studies exist \cite{Shabad,Lammerzahl,causal1}
that take as a starting point consistency conditions such as absence
of superluminal propagation and/or causality, unitarity and stability.
In particular,
in an approach based on the analysis of the photon propagator in a constant
electromagnetic field background \cite{Shabad}, 
the requirements that the group velocity is less than one and that residue
of the propagator be positive (a requirement for unitarity)
can be shown to yield a number of conditions on the first and second
partial derivatives of the Lagrangian density with respect to the
Lorentz invariants $F$ and $G$.
In a completely different approach,
already mentioned in the introduction \cite{Lammerzahl,causal1},
the Fresnel equation for wave propagation in the presence of background fields
was studied.
It was shown that the absence of superluminal propagation implies conditions
on the first and second partial derivatives of the Lagrangian density,
or, equivalently, on the potential $V(P,Q)$,
partially overlapping with the results in \cite{Shabad}.
In this work we took a different philosophy,
analyzing the NLED equations of motion directly for arbitrary potentials,
in particular focusing on singular properties of the equations of motion.
Therefore, it can be regarded as complementary to the earlier studies
\cite{Shabad,Lammerzahl,causal1}.

We showed that a necessary condition for a singularity in the field
equations to occur is that either of the values of $V_P$,
or $S$ as defined in Eq.\ \eqref{S} becomes zero.
While this condition can hold in a large class of Pleba\'nski models,
there are certainly models for which the condition can never be satisfied.
A rather obvious example is Maxwell theory, for which $V_P = -1$ and
$S = 1$ have nonzero constant values.
A less trivial example is Born-Infeld theory:
we see from Eq.\ \eqref{BI_VPQ} that $V_P$ is negative definite,
while the same can be shown to be true for the quantity $S$.
Therefore, for such models the degenerate behavior we studied in this
paper can never occur.

For generic Pleba\'nski models, however, the quantities $V_P$ and/or $S$ 
could become zero for certain ranges of field values.
For the Euler-Heisenberg Lagrangian the onset of degenerate behavior
translates in a breakdown of the model itself,
as it derives from a causal, unitary quantum field theory.
For instance, electron-positron pair production will occur when the
electric field is beyond the Schwinger limit of $1.3\times10^{18}\,$V/m.
This is several orders of magnitude out of reach of even the strongest
laser fields currently available
\footnote{However, experiments are underway to detect QED effects with
colliding petawatt laser pulses (see Ref.\ \cite{lasers}).}.
However, the conditions necessary for reaching singular behavior could well
be relevant in the context of some of the astrophysical or cosmological
scenarios mentioned in section \ref{sec:introduction},
as well as others that have been considered in the literature,
in which Pleba\'nski models have been proposed.

\acknowledgments
R.\ P.\ thanks the kind hospitality of the Center for Applied Space Technology
and Microgravity (ZARM) in Bremen, Germany,
where part of this work was carried out.
C.\ A.\ E.\ was supported by a UNAM-DGAPA postdoctoral fellowship
and the project PAPIIT No.\ IN111518.
R.\ P.\ acknowledges financial support by the Funda\c c\~ao para a Ci\^encia e
a Tecnologia of Portugal (FCT) through grant SFRH/BSAB/150324/2019.


\end{document}